\def\cm{\,{\rm cm}}
\def\Jy{\rm Jy}

\def\cmm2{{\,\rm cm^{-2}}}
\def\cm2{{\,{\rm cm}^2}}
\def\cmm3{{\,{\rm cm}^{-3}}}
\def\gcmm3{{\,{\rm g\,cm^{-3}}}}

\def\m{\hat{m}}

\def\fun#1#2{\lower3.6pt\vbox{\baselineskip0pt\lineskip.9pt
  \ialign{$\mathsurround=0pt#1\hfil##\hfil$\crcr#2\crcr\sim\crcr}}}

\let\grad=\nabla
\def\cross{{\bf \times}}
\def\curl #1 {\grad \cross #1}
\def\div #1 {\grad \cdot #1}

\def\gtwid{\mathrel{\raise.3ex\hbox{$>$\kern-.75em\lower1ex\hbox{$\sim$}}}}
\def\ltwid{\mathrel{\raise.3ex\hbox{$<$\kern-.75em\lower1ex\hbox{$\sim$}}}}

\documentstyle[12pt]{article}
\begin{document}

\thispagestyle{empty}

\begin{center}

{\Large\bf Statistical Tests for the Gaussian Nature \\
of  Primordial Fluctuations Through CBR Experiments}\\

\vspace{.3in}
\vskip 0.2 in
\vspace{.3in}
{\large\bf Xiaochun Luo } \\
\vspace{0.2in}

{ Departments of Physics and of Astronomy \& Astrophysics\\
Enrico Fermi Institute, The University of Chicago, Chicago, IL  60637-1433}\\

{ NASA/Fermilab Astrophysics Center,
Fermi National Accelerator Laboratory, Batavia, IL  60510-0500\footnotemark[1]}

\end{center}

\footnotetext[1]{Address after Oct. 1, 1993 will be: 301 LeConte Hall;
Center for Particle Astrophysics; University of California; Berkeley, CA
94720.}
\vspace{.3in}
\vskip 0.1 in

\centerline{\bf ABSTRACT}
\vspace{0.2in}

Information about the physical  processes
 that generate the primordial fluctuations in the early universe can
be gained by testing the Gaussian nature of the fluctuations through
cosmic microwave background radiation (CBR) temperature anisotropy
experiments. One of the crucial aspects of  density perturbations that are
produced
by the standard inflation scenario is that they are Gaussian,
whereas seeds produced by topological defects left over from
an early cosmic phase transition tend to be non-Gaussian.
To carry out this test, sophisticated statistical tools are required. In this
paper, we will discuss several such statistical tools, including multivariant
skewness and kurtosis, Euler-Poincare characteristics, the three point
temperature correlation function, and the Hotelling's $T^{2}$ statistic defined
through   bispectral estimates of a one dimensional dataset. The effect of
noise present in the current data is discussed in detail and the COBE
53 GHz dataset is analyzed. Our analysis shows that, on the large angular
scale to which COBE is sensitive, the statistics are probably Gaussian.
On the small angular scales, the importance of
Hotelling's $T^{2}$ statistic is stressed,  and the minimum sample size
required to test Gaussianity  is estimated.
Although  the current dataset available from various experiments
 at half-degree scales is still
too small, improvement of the dataset by roughly a factor of two will be
enough to  test the Gaussianity statistically.  On the arcminute scales,
  we analyze the recent RING data
through bispectral analysis, and the result indicates possible
 deviation from Gaussianity.
Effects of   point sources
  are also discussed. It is pointed out that the Gaussianity problem can be
resolved in near future by ground-based or balloon-borne experiments.
\vskip 0.2 in

\vfill\eject

\section{Introduction}
The cosmic structure formation problem is essentially an initial value problem:
how did the universe generate the initial perturbations? In particular,
one can divide the initial condition models into two clear classes: Gaussian
or non-Gaussian? Cosmic inflation \cite{inflation}, on one hand, provides
a natural way to generate Gaussian initial perturbations \cite{structure};
spontaneous symmetry
breaking, on the other hand, will lead to the formation of topological defects
\cite{defect}
via Kibble mechanism \cite{kibble},  and the perturbations generated by
topological defects can be characterized as non-Gaussian. Non-Gaussian
perturbations  also arise in various non-standard inflationary models
\cite{nonGau}.  Thus,
  a test of the Gaussian nature of the primordial perturbations will not only
be helpful in  discriminating different models for structure formation, but
could also shed light on  new physics that yield
 topological defects or special non-standard inflations
 in the early universe.
 Such a test  is, therefore,
 very important and timely in today's cosmology.

  There are two ways to carry out the test. One is from the statistics of the
galaxy counts in a redshift survey \cite{weinberg}. However, since the density
 field we observe today has already gone through the
``black-box'' of non-linear gravitational evolution, one has to  filter
 out this
effect carefully to get a reliable estimate of primordial
quantities \cite{luo1}. In this paper, we will
concentrate on the other approach, which  is from the cosmic microwave
background radiation (CBR) anisotropy experiments. The experiments
 measure the
primeval density perturbations  at redshift $z\sim 1000$.
The density contrast is fairly small at this epoch. The Gaussian nature
of the microwave background fluctuation  directly reflects  the
nature of the primordial perturbations. This approach is promising, especially
after COBE's detection  \cite{cobe} of the   temperature anisotropies at large
angular
scales,
 and the continuing accumulation of data on smaller angular
scales\cite{experiment}.

 Prior to COBE's detection, studies on  CBR  were focused on determining
the
level of anisotropies. The Gaussianity test of the anisotropies was largely
considered as a next-step problem and experimentally intractable. Few detailed
studies \cite{colesb} \cite{string} on the statistics of the CBR
anisotropies have been carried out
except for the Gaussian case \cite{be}. Now that the CBR anisotropies are
detected \cite{cobe}\cite{ganga}, this next-step but important
 question should be
brought into focus and we are optimistic that it can be resolved experimentally
in near future. As we will show later, one doesn't need a  full-sky coverage
at high angular-resolution to study the Gaussianity problem statistically.

There are several recent papers on statistical tests for Gaussianity
\cite{neil}\cite{luo}\cite{moessner}. To carry out the test through
CBR experiments, two important feature have to be stressed. One is the
instrumental noise which  is present in  all current experiments. One should
have a clear understanding of the  instruments and associated noise before
attempting to decipher
non-Gaussian signals from experiments. The other feature is the  smoothing
scale $\theta_{s}$ each experiment operates. Note that the angular size of the
comoving
horizon at decoupling epoch is $ \theta_{c} \sim  2^{\circ}$ by assuming
standard
recombination \cite{kolb}. Put this characteristic scale
 in mind, one can divide
all CBR experiments into three categories: large scales ($\theta_{s}
\gg \theta_{c}$), intermediate scales ($\theta_{s} \sim 1^{\circ}$)
and small scales ($\theta_{s} \sim {\rm arcminutes}$). For large scale
experiments \cite{cobe}\cite{ganga}, each measurement is a sum of anisotropies
in several independent horizons, and one would expect the statistic to be
close to a Gaussian
simply by the virtue of central limit theorem \cite{feller}\cite{scherrer}.
For small scale experiments \cite{meyers}, we will show later that the
data fails the Gaussian statistical tests. However, on arcminute scales,
foreground source contaminations are important. The statistics
of the data may not reflect the statistics of the CBR anisotropies at these
scales. Intermediate scale experiments \cite{interm} are ideal for
 testing Gaussianity. Although the dataset available is still too small,
as we will show in section 4, improvement of the dataset by roughly
a factor of two will be able to test the Gaussinity of CBR statistically.

Several statistical tools are discussed in this paper.
 In section 2, we discuss the simplest tests
 of Gaussianity through the skewness and kurtosis of
the one point distribution.  Skewness and kurtosis are the normalized
third and fourth moments of the distribution and they vanish for Gaussian
distribution.
 Several physically motivated non-Gaussian probability distribution functions
 (PDFs) are considered,   and effects of noise are discussed.
To consider the possible correlation between the
signal and noise, multi-variant skewness and kurtosis are introduced and their
statistics are discussed. In section 3, one geometrical measure of the
random field, the Euler-Poincare (EP) characteristic,
 is discussed and used to
test Gaussianity. The statistics of  the
EP-characteristic and the effects of noise
are discussed and we show that EP-characteristic is hardly a
good discriminator
between Gaussian and non-Gaussian  fields  when the noise
is comparable to the signal.  In section 4, we discuss using the
three point correlation to test Gaussianity. Theoretical
predictions in various models are discussed and we present our analysis
of COBE 53 Ghz data. The result is in good agreement with Gaussian
assumption.   In section 5, we discuss  using Hotelling's $T^{2}$ statistic
to test Gaussianity on intermediate angular scales ($\theta \sim 1^{\circ}$).
The minimum sample size to carry out the test is estimated and sampling
technique are  also discussed. Although the current dataset is still too small
to carry
out the Gaussianity test, improvement of the sample size by roughly a factor
of two will be adequate. 	In section 6, we use the $T^{2}$ statistic
to test RING dataset from OVRO \cite{meyers} on small angular scales
($\sim $ arcminutes). It is found that  the data
  is not consistent with
Gaussian distribution. However, one cannot conclude that the CBR
anisotropies are non-Gaussian on these scales because of the foreground
source contaminations.
  In section 7, we discuss looking for
a special non-Gaussian signal,  the point-like CBR
anisotropy, in small scale CBR experiments.
The Gaussian nature of perturbations from inflation is shown in the appendix.

\section{ Skewness and Kurtosis  of Noisy Data}
The simplest tests of  Gaussianity will be  skewness $\mu_{3}$ and kurtosis
$\mu_{4}$ \cite{luo1}
of the distribution of temperature anisotropies $\delta$,
\begin{equation}
\mu_{3} = m_{3}/\sigma^{3}, \  \ \mu_{4} = m_{4}/\sigma^{4}
-3,
\end{equation}
where $m_{3} $ and $\m_{4}$  are the  third and fourth moments of the
distribution, adn $\sigma$ is the variance of $\delta$. For Gaussian
distribution, both $\mu_{3}$ and $\mu_{4}$ vanish.
In this section, we discuss several physically motivated non-Gaussian
distributions: exponential, log-normal and $\chi^{2}$.
  As we expect,  noise  will blur the effects of non-Gaussian distribution.
The skewness and kurtosis
for these distributions are calculated both with and without noise.
We also discuss the use
of multivariant skewness and kurtosis in cases of noisy data and show how to
estimate them from experimental data.

\subsection{Skewness and Kurtosis of Non-Gaussian Signals}

Coles and Barrow \cite{colesb} have studied the statistics of
a large class of non-Gaussian distributions. We choose the following
distributions based on physical considerations. To reflect the real
experimental setup where the mean of the distribution is subtracted, we
standardize the distribution so that
all of them have a zero mean. Furthermore, we normalize the variance of the
distribution to be unity.  Thus,  distributions we study below
 correspond to the probability distribution functions (PDF) of $x={\Delta
T\over T} \cdot {1\over \sigma_{0}}$, where $\sigma_{0}$ is the observed rms
temperature fluctuations.

\medskip
\noindent
1. Exponential distribution.

This distribution may describe the temperature fluctuation produced
from the cosmic string network on arcminute scales \cite{string}.
The PDF of this distribution is;
\begin{equation}
P (x) = {1\over \sqrt{2}}\exp(- \sqrt{2}|x|) .
\end{equation}
The skewness and kurtosis of the distribution are:
\begin{equation}
\mu_{3} = 0; \  \ \mu_{4} = 1.5.
\end{equation}

\noindent
2. Log-normal distribution.

This distribution is widely used in the statistical studies of galaxies and
 clusters. It is temping to suggest that it might describe the distribution
of temperature anisotropy from Sunyaev-Zeldovich (SZ) effects \cite{sz}.
 Since the effect is produced by the hot-gas in the rich clusters, thus it
should relate intrinsically to the distribution of rich clusters, which is
log-normal.
Simulation of SZ in the
cold dark matter scenario \cite{cole} seems to
support this connection. A small reminder is that the SZ effect always produces
cold
spots on the sky;   thus,  the distribution of the temperature fluctuation is
different from the usual log-normal
distribution by a sign. The PDF is given by:
\begin{equation}
P(x) = {1\over \sqrt{2\pi}\sigma (-x)} \exp( -(log (-x))^{2}/\sigma_{2}), x<0.
\end{equation}
where $\sigma$ is given by:
\begin{equation}
\sigma = {1\over 2} \ln {1 + \sqrt{5}\over 2},
\end{equation}
The skewness and kurtosis of the distribution are:
\begin{eqnarray}
\mu_{3} =- {\exp(3\alpha) - 3\exp(\alpha) + 2 \over (\exp(\alpha) -1)^{3/2}}
\nonumber \\   \mu_{4} = {\exp(6\alpha) - 4 \exp(3 \alpha) + 6 \exp(\alpha) -3
\over (\exp(\alpha) -1)^{2}},
\label{log-n}
\end{eqnarray}
where $\alpha = \sigma^{2}.$
Thus,
\begin{equation}
\mu_{3} = -0.66,  \mu_{4} = 2.72.
\end{equation}

\noindent
3. $\chi_{n}^{2}$ distribution.

This class of distributions
 provides a good fit to the statistics of temperature
fluctuation from global topological defects and non-topological defects in
the framework of $O(N) \ \  \sigma$-model \cite{sigma}.
In this model, a global symmetry $O(N)$ is broken to $O(N-1)$ by a N-component
real scalar field $\phi = (\phi_{1}, ..., \phi_{N})$ in the early universe.
The temperature anisotropy produced by the dynamics of the scalar fields is
given by:
\begin{equation}
{\delta T \over T} = - {2\pi G \eta^{2} \over 9} \Theta_{00} =- {2\pi G
\eta^{2} \over 9} \sum_{i} [\dot{\phi_{i}}^{2} + (\nabla \phi_{i})^{2}].
\label{turok}
\end{equation}
When $N$ is small, the dynamics of the scalar fields are nonlinear, thus, the
PDF for $\phi_{i}$ is non-Gaussian. However, when $N$ is larger, the dynamical
equations for $\phi_{i}$ decoupled and become linear \cite{sigma}
 and therefore $\phi_{i} ( i =1, ..., n) $ become independent
Gaussians. From Eq. (\ref{turok}), it is clear that in the large N limit, the
temperature anisotropy is $\chi^{2}$ distributed with $n=4N$ degrees of
freedom.
  Because of our standardization process, the PDF is  related to the
usual $\chi^{2}$ distribution by the following transformation:
\begin{equation}
x \rightarrow x + \sqrt{n/2}, \  \ \sigma^{2} \rightarrow 1/\sqrt{2n}.
\end{equation}
Thus, the PDF of the distribution is given by:
\begin{equation}
P(x) = {1\over (2/n)^{n/2} \Gamma (n/2) }  (x + \sqrt{n/2})^{{n-2\over2}} \exp
(-{{\sqrt{2n}\cdot x + n} \over 2}), x \geq -\sqrt {n/2}.
\label{chi}
\end{equation}
This PDF is plotted in Fig. (1) for $n=4, 8, 16$. The distribution
is very non-Gaussian for low $n$ but approaching asymptotically toward Gaussian
in the large $n$ limit. The skewness and kurtosis of the distribution of this
class of distribution is:
\begin{equation}
\mu_{3} = \sqrt{8/n}, \  \ \mu_{4} = 15/n.
\end{equation}

With the presence of Gaussian  noise, the skewness and kurtosis will reduce
dramatically. While the third and fourth moments of the distribution are
unchanged,
the variance increases by $ (1 + \alpha^{2})$, when the signal to noise ratio
is $1:\alpha$. Thus,
 the skewness will reduce by $ (1 + \alpha^{2})^{-3/2}$  and
the kurtosis by $ (1 + \alpha^{2})^{2}$. For the log-normal distribution, the
relations mentioned above are not true. The skewness and kurtosis can be
 calculated from Eq.(\ref{log-n}).  The skewness and kurtosis for various
non-Gaussian distributions are listed in Table 1. In the noisy case, the signal
to noise ratio is set to be 1:1.

\subsection{Estimate of Skewness and Kurtosis of Noisy Data}

  In this section, we will address the following questions:
how to estimate the skewness and kurtosis from noisy experimental
data,  and what are the statistics of these quantities when the signal and
noise are both Gaussian?

In the case where the experimental data is noisy,  one has to deal with
 two  random  variables, the signal and the noise, which have a bivariant
joint distribution. Thus, we have to generalize the usual skewness and
kurtosis to bivariant distributions.
 Let us first consider the general  multivariant distribution.
For a $p-$dimensional random vector $\bar{X}=(x_{1}, x_{2},
..., x_{p})$ with zero mean and covariance matrix
$\Sigma$,  it is helpful to introduce
the following multivariant measures of skewness and kurtosis \cite{rao}:
For  skewness,
\begin{equation}
\beta_{1p} = \sum_{r,s,t=1}^{p}\sum_{r^{'},s^{'},t^{'}=1}^{p}
\sigma^{rr^{'}}\sigma^{ss^{'}}\sigma^{tt^{'}}\mu_{111}^{rst}
\mu_{111}^{r^{'}s^{'}t^{'}},
\end{equation}
where $\mu_{111}^{rst}=<x_{r}x_{s} x_{t}>$, $\sigma^{ij}$ is the $i, j$th
element of $\Sigma^{-1}$, the inverse of covariant matrix $\Sigma$.
If the joint-distribution function for $\bar{X}$ is a multivariant Gaussian,
then $\beta_{1p}=0$.  $\beta_{1p}$
can be estimated from a sample of size $N$, where we can replace the
ensemble-averaged $\mu_{111}$ and covariance matrix elements with the
sample-averaged ones\footnote{Here, we explicitly assume that the temperature
fluctuation
is ergodic so that the ensemble average will be identical to the spatial
average
(in our case, it is the sample average) in the limit of large spatial coverage.
The ergodicity is guaranteed if the power spectrum of the fluctuation is
continuous. See Adler \cite{adler} and Bardeen, Bond, Kaiser \& Szalay
\cite{bbks}
 for details.}.
 If we denote the estimate of $\beta_{1p}$ by $b_{1p}$, then under the Gaussian
hypothesis, the statistic $A={N b_{1p}\over 6}$ is approximated distributed
as a $\chi^{2}$ with ${p(p+1)(p+2)\over 6}$ degrees of freedom.
 For us, the interesting cases are for $p=1, 2$. When $p=1$,
$\beta_{11}=\mu_{3}$, the usual skewness. If one denote $b_{3}$ as the estimate
of skewness $\mu_{3}$ from a sample of size $N$, then
\begin{equation}
b_{3} = {1\over N} \sum_{i=1}^{N} \delta_{i}^{3},
\end{equation}
and $ A={N b_{3}\over 6}$ is distributed as a $\chi^{2}$ with 1 degree of
freedom.
For $p=2$,
\begin{eqnarray}
\beta_{12} = (1-\rho^{2})^{-3} \{\gamma_{30}^{2} + \gamma_{03}^{2}
+ 3(1 +2\rho^{2})(\gamma_{12}^{2} + \gamma_{21}^{2})
-2\rho^{3}\gamma_{30}\gamma_{03}
 \nonumber \\+ 6\rho [(\gamma_{30}\rho\gamma_{12}
-\gamma_{21})
 + \gamma_{03}(\rho\gamma_{21}-\gamma_{12}) -(2+\rho^{2})
\gamma_{12}\gamma_{21}]\},
\end{eqnarray}
where
\begin{equation}
\rho={M_{12}\over \sigma_{1}\sigma_{2}}, \gamma_{rs}={\mu_{rs}\over
\sigma_{1}^{r} \sigma_{2}^{s}}, \mu_{rs}=<x_{1}^{r} x_{2}^{s}>.
\label{grs}
\end{equation}
In the case where signal and noise are both random Gaussian fields, the
joint probability distribution  function is a bi-variant
Gaussian:
\begin{equation}
P(x_{1}, x_{2}) = {1\over (2\pi)^{2} det M} \exp(- 1/2 \sum x_{i} M_{ij}^{-1}
 x_{j}),
\end{equation}
where the correlation matrix between the signal and noise is the following:
\begin {equation}
M_{11} = \sigma_{1}^{2}, M_{22} = \sigma_{2}^{2},
M_{12} = M_{21} = < x_{1} x_{2}>.
\end{equation}
For COBE DMR, the correlation matrix can be estimated from (A+B)/2 and
(A-B)/2 maps. The bi-variant measure of skewness $\beta_{12} =0$ if we assume
 that the signal and noise are both Gaussian. Let us  denote $b_{12}$
as the estimate of $\beta_{12}$ from a sample of size $N$ (say, COBE map).
Under the
Gaussian hypothesis, $A={N b_{12}\over 6}$ is distributed as a $\chi^{2}$
with 4 degrees of freedom.

For kurtosis, we have the following measure:
\begin{equation}
\beta_{2p} = [<\bar{X} \Sigma^{-1} \bar{X}>]^{2} - p(p+2).
\end{equation}
 Given a random sample of size N on the random vector $\bar{X}$, we can
replace the ensemble average with the sample average and  estimate
$\beta_{22}$ by
\begin{equation}
b_{2p} = {1\over N} \sum_{1}^{N} [(\bar{X}_{i} -\mu) S^{-1} (\bar{X}_{i}
-\mu)]^{2} -p(p+2),
\end{equation}
where $\mu = {{1\over N} \sum_{1}^{N} \bar{X}_{i}}, $ and
$S_{ij} = {1\over N} \sum_{1}^{N}(\bar{X}_{i} -\mu)(\bar{X}_{j}-\mu)$.
For a multivariant Gaussian, $b_{2p} =0$,  and the following statistic
$B = [{b_{2p} \over {8 p (p+2) \over N}}]^{1/2}$ is approximately distributed
as normal with zero mean and variance one.

For p=1, $\beta_{21}$ reduce to $\mu_{4}$, the usual
kurtosis we defined before. An estimate of $\mu_{4}$ from a given sample
is
\begin{equation}
b_{4} = {1\over N} \sum_{i = 1}^{N} x_{i}^{4}.
\end{equation}
and the sample variance is $\sqrt{24/N}$.

For p =2,
\begin{equation}
\beta_{22} =-8 + \{ \gamma_{40} + \gamma_{04} +
4 \rho (\rho\gamma_{22} - \gamma_{13}\gamma_{31})\}/ (1-\rho^{2})^{2},
\end{equation}
where $\rho$ and $\gamma_{rs}$ is the same as the ones defined in Eq. (
\ref{grs}). Under Gaussian assumption,
$\beta_{22} = 0$, and the estimation of $\beta_{22}$, $b_{22}$ has zero mean
and sample variance $ \sqrt{48/N}$.

\subsection{ COBE data}
At present, the only existing complete dataset of temperature anisotropies
is the COBE dataset. Thus, it is tempting to ultilize the statistics we
discussed before and put a constraint on the possible deviations from
Gaussianity
 through
COBE data. Unfortunately,  the skewness and kurtosis
of the distribution of $\Delta T\over T$ cannot be estimated directly from the
COBE dataset because the signal in each pixel is not an independent measurement
of  $\Delta T\over T$. Furthermore, in order to estimate skewness and kurtosis,
we have to assume that the temperature anisotropies are ergodic so that the
spatial average is equivalent to the ensemble average. But, even for Gaussian
fluctuations, the ergodic hypothesis is true only if the power spectrum of the
fluctuations is continuous \cite{adler}\cite{bbks}. The observed temperature
anisotropies  are 2D random fields on a 2-sphere. The power spectra
$C_{l}$ is discretized and asymptotically approaching continuity in the large
$l$ limit. For the COBE experiment, which  is sensitive only to the low
$l$ moments, the ergodic hypothesis doesn't hold. Thus,
the  statistics estimated from COBE directly will only
be  the estimates of the ``local'' values: they are the measure
of deviation from Gaussiantiy in our horizon. To estimate
the cosmic skewness and kurtosis, i.e., the skewness and kurtosis
averaged over an ensemble of horizons, one
 has to treat the observed value as a N-dimensional
random vector,  where N is the sample size (the total number of pixels).
Since we have only one measurement (we have  only one universe),
in order
to calculate the ensemble average of the quantities
equation, the only conceivable way is to use
Monte-Carlo simulation. The COBE analysis along this line will soon appear
\cite{smoot} and  won't be repeated here.
  We note  that by using the statistics
of multivariant skewness, which is a $\chi^{2}$ distribution
with $N (N+1) (N+2)/6$ degrees of freedom.
The  total number of simulated maps must exceed
$N_{map} \ge {N\over \beta_{3}^{2}}$,
where $\beta_{3} $ is the variance of the skewness of a distribution.
 To estimate
a skewness with variance smaller than  0.1, the number of simulated
maps has to be larger than
$ 6 \times 10^{5}$.

\section{Topological Measures of a Random Field}
In this section, we will discuss the topological measure of a
random field and the application to the test of deviation
 from Gaussianity. This approach was
  studied in \cite{coles}\cite{gott}.
  It is found that among all these quantities,
the Euler-Poincare (EP)
 characteristic is the  most effective topological measure
with regard to  testing Gaussianity.
Adler \cite{adler} also derived the mean for a special non-Gaussian field: the
$\chi^{2}$ field.  Subsequently, Coles \cite{coles} applied the result to
a number of non-Gaussian fields which are derived from Gaussian. Gott
et al. \cite{gott} applied EP characteristic to simulated cosmic string maps.
Both confirmed that the EP  characteristic is effective in testing Gaussianity.
We will briefly review the existing results
obtained by previous investigators, then we will move on to study the
statistics
of the topological quantities. Special attention is paid to the real
 experimental
situation where the noise level is high. We also derive some new results
on  topological measures in the presence of Gaussian noise.
 We will show that in the case
where the signal to noise ratio is around 1:1, the EP characteristic,
unfortunately,
fails to be effective in  discriminating
 between Gaussian and non-Gaussian distributions.
\subsection{Mean }
The central concept on which all topological measures are based
is the excersion set \cite{adler} which is the set of points where the field
$F(r)$ exceeds a global
value $u$. If we take a map  of a certain area of the CBR sky, the excersion
set of the map above level $u$, denoted by $S_{u}$, will in general consist of
a
number of disjoint regions, each having a boundary which is the contour of
 satisfying $ F(r) = u$. Since the
 smoothness and the differentiability of the contours are
guaranteed by the beam smoothing\footnote{For noise, without smoothing, the
contour will be a fractal and it will not be differentiable. We illustrate
this point in Fig. 2. For a detailed discussion, see cite{adler, coles}.}, the
Euler-Poincare (EP) characteristic of $S_{u}$, $\Gamma_{u}$ is given by:
\begin{equation}
\Gamma_{u} = {1\over 2\pi} \int k ds,
\end{equation}
where $k$ is the geodesic curvature of the contours and the intergal is taken
over all contours in the map.

For Gaussian random fields, the mean of the
EP characteristic is exactly calculatable:
\begin{equation}
<\Gamma_{\nu} > = {1\over (2\pi)^{3/2}} ({C_{2}\over C_{0}})^{2} \nu
\exp(-\nu^{2}/2),
\end{equation}
 where $C_{0} = <T^{2} (\hat{r})>$,
$C_{2} = <~\nabla~T(\hat r) \nabla T(\hat r)~>$ is the variance of
 the temperature anisotropy gradient\footnote{$C_{2}/C_{0}$ is equal
to $\sqrt{2} \gamma^{2}/\theta_{*}^{2}$, where $\gamma$ and $\theta_{*}$
is the spectral parameters defined by \cite{be}.}, and $\nu = u/\sigma$,
$\sigma$ is the rms temperature anisotropy.

Among all possible non-Gaussian PDFs, we specifically choose the $\chi^{2}_{n}$
distribution because of its relevance to the $O(N) \ \ \sigma$-model. For our
standardized PDF given in Eq. (\ref {chi}), the mean number of the EP
characteristic
is given by:
\begin{equation}
<\Gamma_{\nu}> = {1\over 4\pi \Gamma (n/2)} ({C_{2}\over C_{0}})^{2}
({\sqrt{n} \nu  + n \over 2})^{(n-2)/2} ( \sqrt{n} \nu +1) \exp (-
(\sqrt{n} \nu + n)/2).
\end{equation}

\subsection{Statistics}
The Euler-Poincare characteristic is  a discrete point process defined
over the underlying random field. For any point process $N(\hat{r})$ with
mean $<N> = \bar{N}$, the variance of the process is given by \cite{peebles}:
\begin{equation}
<N^{2}> = |\bar{N}| + \bar{N}^{2} ( 1 + \bar \xi),
\label {discrete}
\end{equation}
where
\begin{equation}
\bar{\xi} = {1\over A^{2}} \int d\Omega_{1} d\Omega_{2} \xi (\hat r_{12})
\end{equation}
is the sample-area averaged two point correlation of the process $N$.
The first term in Eq. (\ref {discrete}) is due to the discreteness of the
process. The variance of the process is given by:
\begin{equation}
\sigma^{2} = <N^{2}> - \bar{N}^{2} = \bar{N} + \bar{N}^{2} \bar{\xi}.
\end{equation}
  Thus, to find the variance of the EP characteristics,
the central problem is to find the  two point correlation.
We approximate this correlation function $\xi$ by the peak-peak correlation
$ \chi$ of the same underlying random field. Since $\Gamma (\nu)$ is defined
through
the $\nu$-peaks of the density field, the approximation should provide
a reasonable fit to the true $\Gamma$ correlation.
We expect that this approximation will give poor results when $\Gamma (\nu)
\sim
0$.

The peak-peak correlation function is well studied in the
literature \cite{peak} \cite{be}
and is most easily understood as the following:
Let $P_{1} (\hat{r})$ be the probability of finding one $\nu-$peak at $\hat r$
 and $P_{2}$ be the probability of finding a $\nu-$ peak at $\hat r_{1} $
and $\hat r_{2}$. Then, the two point correlation for the peaks will be:
\begin{equation}
1 + C_{2} (\hat{r}_{1}, \hat{r}_{2}) = P_{2}/P_{1}^{2}.
\end{equation}
For a Gaussian random field, we have
\begin{equation}
P_{1} = {1\over \sqrt{2\pi} \sigma_{0}} \int_{\nu\sigma_{0}}^{\infty} dx
\exp (- {x^{2} \over 2 \sigma_{0}^{2}}),
\end{equation}
and
\begin{equation}
P_{2} = {1\over 2\pi {\rm det} M} \int_{\nu\sigma_{0}}^{\infty}
\int_{\nu\sigma_{0}}^{\infty} dx_{1} dx_{2}\exp( - {1\over 2} \bar{\bf x}
M^{-1} {\bf{x}}),
\end{equation}
where $\bf x = (x_{1}, x_{2})$ and $M (r_{1}, r_{2})$ is the correlation
 matrix of the random  field at $r_{1},  r_{2}$, which is
\begin{equation}
M = \left|\begin{array}{cc} 1 & \psi (\theta) \\
\psi (\theta) & 1
\end{array}\right| \cdot \sigma_{0}\nonumber
\end{equation}
 and $\psi (\theta)$ is the normalized two point function.
Thus, in the limit where $\psi \ll 1$, the peak-peak correlation function is
given by:
\begin{equation}
\xi_{peak} = \nu^{2} \psi (\hat{r}).
\end{equation}
The beam-smoothed two point temperature correlation is well approximated by
:
\begin{equation}
C(\theta) = C(0) \exp (- \theta^{2}/\theta_{c}^{2}),
\end{equation}
where the coherent angle $\theta_{c} \approx \theta_{*}$,
is a function of the beam width.  The averaged correlation in an area
$A = \pi \theta_{A}^{2}$ is
\begin{equation}
\bar{\psi} =  \nu^{2} ({\theta_{c}\over \theta_{A}})^{2}.
\end{equation}
Thus, the variance of the Euler-Poincare characteristic is given by:
\begin{equation}
\sigma^{2} = |\bar{N}| + \nu^{2} ({\theta_{c}\over \theta_{A}})^{2}
\bar{N}^{2},
 \bar{N} = n (\pi \theta_{A}^{2}).
\label{ep}
\end{equation}
The mean and $1 \sigma$ uncertainty of the EP characteristic
$\Gamma (\nu)$ for Gaussian random field is plotted in Fig. 3. Because of
the approximations we used to derive the statistic, we
expect the uncertainties of $\Gamma (\nu)$ given by Eq. (\ref{ep})
are not exact when $\nu \sim 0$.

\subsection{The EP Characteristic of Noisy Data}
Our major concern over the applicability of the EP characteristic as
a reliable statistic to discriminate between a Gaussian and a
non-Gaussian random field is the noise term, which appears in all current CBR
experiments and is comparable in amplitude to the signal. In this section, we
will show first
 how the beam used in the experiments
regulates the noise. Without smoothing,  noise will be an obvious
hazard to
 topological measures  because the rms of the
derivatives  of noise
is not well defined.  We will then proceed to study the change of  EP
characteristic
due to the beam-smoothed  noise term.

The spectral parameter of the noise is given by:
\begin{equation}
\sigma_{0}^{2} = A \sum_{l=2} (2 l +1) \exp( - l(l+1) \sigma_{s}^{2})
\sim A ({1\over \sigma_{s}})^{2},
\end{equation}
\begin{equation}
\sigma_{1}^{2} = A \sum_{l=2} l (l+1) (2 l +1) \exp( - l(l+1) \sigma_{s}^{2})
\sim {1\over 2} A ({1\over \sigma_{s}})^{4}
\end{equation}
\begin{equation}
\sigma_{2}^{2} = A \sum_{l=2} (l-1) l (l+1) (l+2) (2 l +1) \exp( - l(l+1)
\sigma_{s}^{2}) \sim {1\over 3} A ({1\over \sigma_{s}})^{6}.
\end{equation}
Thus, $\gamma = {\sigma_{1}^{2} \over \sigma_{0} \sigma_{2}} = 0.75$ and
 $\theta_{*} = \sqrt{2} {\sigma_{1}\over \sigma_{2}} =
\sqrt{3} \sigma_{s}$. If we compare
this value for noise with that of a Harrison-Zeldovich primordial perturbation,
where $\gamma \sim 0.5$ and $\theta_{*} \sim 1.8 \sigma$, we can conclude that
the noise term regulated fairly well by the beam.

However, even with a beam-smoothing, the EP characteristic of the observed
random temperature anisotropy pattern is still
changed due the presence of noise.  We first consider a situation where
the sky is dominated by a quadrupole only, as we showed Fig. (4a).
The EP
characteristic is very simple for this map: $\Gamma =2 \ \
{\rm for} \ \ \nu = -1, 1, 2.$ However, when one puts Gaussian noise of the
same
variance to the signal, the map is dominated by the feature of noise,
 which are shown in Fig. (4b).
Even though we don't expect the temperature anisotropies to be a pure
quadrupole,
this example makes us cautious when  using the EP characteristic.
In the  realistic cases where
the signal itself is also a random field which is non-Gaussian, the problem is
harder.
A full analysis of the EP characteristic
for the sum of two general random fields is not tractable. Only one
special case where both fields are Gaussian is solved \cite{adler}.
 In our problem,  where one field is non-Gaussian, there is no  ready-to-use
result to apply. Thus, we solve this problem by the following strategy:
the $\chi_{n}^{2}, n =1, ...$ is a class of distributions, ranging from
very non-Gaussian (small n) to slightly deviating from Gaussian (large n), and
the EP characteristic of this class of distribution is known. Thus, we first
find the modified non-Gaussian PDF due to the presence of noise, then we
find the best fit $\chi^{2}$ distribution to this modified PDF. The EP
characteristic of the noisy temperature map is thus approximated to be
the EP characteristic of the best fit $\chi^{2}$ distribution.
The validity of this approach lies partly in the fact that if the difference
between
two PDFs goes to zero, the difference between the corresponding EP
characteristic also goes to zero.

We assume the noise is Gaussian and uncorrelated with the CBR anisotropy
signals. The PDF for the noise is:
\begin{equation}
P (\eta) = {1\over \sqrt{2\pi}  \sigma_{n}} \exp( -\eta^{2}/2\sigma_{n}^{2})
\end{equation}
where $\sigma_{n}$ is the variance of noise. In the following, we consider
only the case where the signal to noise ratio is 1:1, thus $\sigma_{noise}
=\sigma_{signal} = 1/\sqrt{2}$.  It is straightforward
to generalize to the arbitrary noise case, and we will show it here.

The PDF for the noisy map is given by:
\begin{equation}
P(x) = \int P_{\rm signal}(x^{'}) P_{\rm noise} (x-x^{'}) dx^{'}
\end{equation}
It is convenient to use the cumulant function \cite{cum} $K(u)$, which is the
logarithm of the
characteristic function $\phi (u)$,  the Fourier transformation of
the PDF:
\begin{equation}
\phi (u) = \int P(x) e^{i u x} dx.
\end{equation}
The cumulant function for $P(x)$ is simply the sum of the cumulant
 function of the signal and the
noise. For Gaussian noise, the cumulant function is
simple: $K_{noise} (u) = - {u^{2}\over 4}$. For the signal, the PDF is the
modified $\chi^{2}$ distribution. After some algebra, the cumulant is found to
be:
\begin{equation}
K_{signal} (u) = {n\over 2} \ln ( 1- {i u\over \sqrt{n}})
 + i \sqrt{n} u/2.
\end{equation}
Thus, the cumulant function for the noisy signal is given by:
\begin{equation}
K (u ) = K_{noise} (u) + K_{signal} (u)
 \approx - u^{2}/2 + {i\over 6 \sqrt{n}} ({u\over \sqrt{2}})^{3}, \  \ {\rm n
\gg 1}
\end{equation}
Thus, with noise, the cumulant is still that of a $\chi^{2}$ distribution
in the limit of large $n$, but
with $N$ degrees of freedom, where $N = 2^{3/4} n \sim 1.7 n$.
In Fig. 5, we plot the EP characteristic for $n= 12$ with and without
noise. Even for the global monopole where $n \simeq 12$,
the EP characteristic fails to be effective.

\section{Three Point Temperature Correlation Function }
 There are various examples where  non-Gaussian
processes  possess Gaussian PDF \cite{feller}. One classic example is the
smoothed Poisson
point process. The process is non-Gaussian when the smoothing scale is small,
and  tends to be Gaussian when the smoothing scale becomes large
by the virtue of the central limit theorem \cite{scherrer}.
Thus, the test of Gaussianity should go
beyond the mere one point PDF. The EP characteristic discussed in the previous
section is one way to take into account the full properties of a random
Gaussian
field. But the noise present in the data prevents it from being effective.
However, as we stressed before \cite{luo}, the
three point temperature correlation function is a good measure of deviations
from Gaussianity for the noisy data, as long as the noise is mutually
independent
and not correlated with signals. In this section, we will first
introduce statistics to test these aspects of noise. Then, we will
discuss theoretical predictions of the three point function in different models
\cite{three}. The reduced three-point functions for COBE 53 GHz signal and
noise maps are obtained and show no deviation from Gaussian.
We conclude that at the COBE scale, the temperature anisotropies are probably
Gaussian.

\subsection{Properties of Noise}
A good understanding of noise in CBR experiments is crucial
in testing the Gaussianity of the primordial density perturbation
through the existing data.
One has to make specific assumptions about the instrumental noise
in order to test the Gaussianity: the instrumental noise must be mutually
independent among pixels
and Gaussian. The assumption has to be tested thoroughly
before any attempt to decipher the non-Gaussian CBR signal from the data.
The statistical tools of testing Gaussianity we introduced throughout
this
 paper should also apply to the noise and we wouldn't repeat them
here. In this
section, we will address the following aspects related to the noise
 in the experimental data: (1) is the noise mutually independent? (2)
does noise correlate with the CBR signals?

In testing the mutual independence of the noise, we use the following results
from
statistics \cite{rao}: Let {$y_{i}, (i =1, ..., N)$} be a measurement of
 a zero mean random process. The mean and two point correlation of the
process can be estimated as the following:
\begin{eqnarray}
\bar{y} = {1\over N} \sum_{i=1}^{N} y_{i}, \\
C_{2} (s) = {1\over N} \sum_{i=1}^{N- |s|} (y_{i} -\bar{y}) (y_{i +|s|}-
\bar{y}), s =0, \pm1; ..., \pm (N-1),\\
\rho (s) = C_{2} (s) / C_{2} (0).
\end{eqnarray}
Then, if $y_{t}$ are mutually independent, then $\rho (s)$ is asymptotically
Gaussian. In particular,
 $\rho (1)$ is asymptotically Gaussian with zero mean and
variance ${1\over N}$.  With
 a suitable
redefinition of ${y_{i}}$, this result can be applied to answer questions
 (1) and (2).

\medskip
\noindent
{\bf (1)Testing for the Independence of the Noise}

Let $\eta_{i}, i =1, ..., N$, be the noise in $i$th pixel. The mutual
independence of noise can be tested through a second covariance of analysis of
the square of the noise. Let
$y_{i} = \eta_{i}^{2}, i =1, ..., N$. If $\eta_{i}$ are not correlated,
then  $y_{i}$ will not be either. Thus,  we can define the following statistic:
\begin{equation}
W_{n} = {\rho(1) C_{2}(0) \sqrt{N} \over \sqrt{C_{4} (0, 1, 1)}},
\end{equation}
where
\begin{equation}
C_{4}(0, 1, 1) = {1\over N} \sum_{i =1}^{N-1} (y_{i} - \bar{y})^{2} (y_{i+1}
-\bar{y})^{2}.
\end{equation}
Under the hypothesis that the noise are independent,
$W_{n}$ is distributed as a
standard Gaussian (zero mean and unity variance). We
can define $W_{s}$, the same statistic as $W_{n}$ by using the signal in each
pixel, to test the independence of the signal.

\medskip
\noindent
{\bf (2) Testing the Correlation among Signal and Noise}

Noise should not only be uncorrelated, but should also be independent
of the signal. We assume that the noise is additive, i.e.,
$\delta_{obs} = \delta_{CBR} + \eta$. Once the noise is found to be
independent, the variable
\begin{equation}
y_{i} = \eta_{i}^{2} \delta_{i}^{obs} = \eta_{i} ^{2}
(\delta_{i} + \eta_{i}), \nonumber
\end{equation}
should also be mutually independent.
Following the previous section, the following set of statistics
\begin{equation}
W_{c}  = {\rho(1) C_{2}(0) \sqrt{N} \over \sqrt{C_{4} (0, 1, 1)}},
\end{equation}
with
\begin{equation}
C_{4}(0, 1, 1) = {1\over N} \sum_{i =1}^{N-1} (y_{i} - \bar{y})^{2} (y_{i+1}
-\bar{y})^{2},
\end{equation}
is a standard Gaussian.

\medskip
\noindent
{\bf (3) Analysis of COBE GHz DMR Map}

For COBE DMR, the (A+B)/2 is  the signal+noise map and
the (A-B)/2 is the noise map. The properties of signal and
noise are tested through the statistics defined aboved and
shown in Table 2.

The analysis shows  that at the $1.6\%$
confidence level, the noise is uncorrelated. At the $19\%$ confidence
level, the noise is not correlated with the signal. Thus, we can conclude that
the
noise in COBE 53 GHz map  is not correlated with the signal and
marginally uncorrelated.
The $W_{s}$ in the last row is the statistic to test the
independence of the signal. The statistic shows that the
hypothesis that the data is not correlated  failed badly.
This is expected if the signal is primordial CBR fluctuations.

\subsection{Three Point Function}
It is convenient to use the normalized two point $\psi$ and three
point function $\eta$, where
\begin{equation}
\psi (|\hat{r}_{1} - \hat{r}_{2}|) = C_{2} (\hat{r}_{1}, \hat{r}_{2})/C_{2}(0),
\  \ \psi (0) = 1
\end{equation}
\begin{equation}
\eta (\hat{r}_{1}, \hat{r}_{2}, \hat{r}_{3}) = <{\delta T \over T}
(\hat{r}_{1})
{\delta T \over T} (\hat{r}_{2}) {\delta T \over T} (\hat{r}_{3})> /C_{2}
(0)^{1.5}, \  \ \eta(0) = \mu_{3},
\end{equation}
where $\mu_{3}$ is the skewness.
The theoretical predictions for three point function are mainly the following:

\medskip
\noindent
{\bf (1) Inflation}

Various non-standard inflation models \cite{nonGau}
will generate a non-zero three point correlation function. The generic form of
the three point function in most inflationary models is:
\begin{eqnarray}
\eta (\hat{r}_{1}, \hat{r}_{2}, \hat{r}_{3}) = {\lambda\over 3}
(\psi (|\hat{r}_{1} - \hat{r}_{2}|) \psi (|\hat{r}_{1} - \hat{r}_{3}|) +
\nonumber
\\
\psi (|\hat{r}_{1} - \hat{r}_{2}|) \psi (|\hat{r}_{2} - \hat{r}_{3}|) +
\psi (|\hat{r}_{1} - \hat{r}_{3}|) \psi (|\hat{r}_{2} - \hat{r}_{3}|)),
\end{eqnarray}
where $\lambda$ is a dimensionless constant. As we will show in the Appendix,
for one field slow-roll inflation models, $\lambda \sim 10^{-7}$. In
non-standard inflation models \cite{nonGau}, $\sigma$ can be much larger
(up to order unity).

\medskip
\noindent
{\bf (2) $\chi^{2}$ fields.}

Consider the $\chi^{2}$ field $ Y = \sum_{i=1}^{n} X_{i}^{2}$.
As we discussed before, the $\chi^{2}$ field describes the global topological
defects in the large $ N$ limit. By extrapolating to low N, one can also
gain insights into the possible temperature anisotropy patterns generated by
domain walls,  strings, monopoles or textures.  The two point
correlation is given by:
\begin{equation}
C_{2} (\hat{r}_{1}, \hat{r}_{2}) = 2n \phi^{2} (\hat{r}_{1}, \hat{r}_{2}),
\end{equation}
where $\phi$ is the common covariance function of the $X_{i}$. We choose $\phi$
so that the two point function of $Y$ matchs the observation,
\begin{equation}
\phi^{2} (\hat{r}_{1}, \hat{r}_{2}) = C_{2} (\hat{r}_{1}, \hat{r}_{2})/2n.
\nonumber
\end{equation}
The three point function is found to be:
\begin{equation}
\eta (\hat{r}_{1}, \hat{r}_{2}, \hat{r}_{3}) = \sqrt{8/n} (\psi (|\hat{r}_{1} -
\hat{r}_{2}|) \psi (|\hat{r}_{1} - \hat{r}_{3}|) \psi (|\hat{r}_{2} -
\hat{r}_{3}|))^{3/2}
\end{equation}

\noindent
{\bf (3) Late-time Phase Transition}

In this model \cite{hill}, due to the conformal invariance of the system at the
critical point \cite{polyakov}, the three
point function result from this class of cosmological phase transitions has the
following simple form:
\begin{equation}
\eta (\hat{r}_{1}, \hat{r}_{2}, \hat{r}_{3}) = \alpha (\psi (|\hat{r}_{1} -
\hat{r}_{2}|) \psi (|\hat{r}_{1} - \hat{r}_{3}|) \psi (|\hat{r}_{2} -
\hat{r}_{3}|))^{3},
\end{equation}
where $\alpha$ is a dimensionless constant of order unity.

The full structure of $\eta$ is complicated. We consider the reduced three
point
function\cite{reduce} where
\begin{equation}
\eta_{reduced} (\hat{r}_{1}, \hat{r}_{2}) =  <({\delta T \over
T})^{2}(\hat{r}_{1})  {\delta T \over T} (\hat{r}_{2})>/ C_{2} (0)^{3/2}.
\end{equation}
The theoretical predictions in various models are given by:
\begin{equation}
\eta_{reduced} (\hat{r}_{1}, \hat{r}_{2}) = {\mu_{3}\over 3} ( 1 + 2 \psi
(|\hat{r}_{1} - \hat{r}_{2}|)) \ \ {\rm for \ \ inflationary \ \ models}
\end{equation}
\begin{equation}
\eta_{reduced} = \sqrt{2/N} \psi^{3} (|\hat{r}_{1} - \hat{r}_{2}|)
 \  \ {\rm  for \ \ O(N) \sigma - models}
\end{equation}
\begin{equation}
\eta_{reduced} = \alpha \psi^{6} (|\hat{r}_{1} - \hat{r}_{2}|)
 \  \ {\rm  for \ \ LTPT \ \  models}
\end{equation}
The three point functions from the previous three different categories
are plotted in Fig. 6,  the skewness is chosen to be the same for all
cases.

\subsection{Analysis of COBE data}
The   53 GHz COBE DMR data  is analyzed ultilizing the statistics we discussed
above. This frequency (53 GHz) is chosen because it has the best data quality
\cite{cobe}. Before subtracting the dipole and beginning
further analysis, the signal is weighted
by the estimated pixel uncertainty. The dipole is subtracted by using the
most recent COBE result \cite{dipole}. For 53 GHz (A+B)/2 map, the
subtracted dipole signal is:
\begin{equation}
T(l, b) (mK) = -0.198 \cdot \cos (l)\cos(b)
-2.075 \cdot \sin(l) \cos(b) + 2.333 \cdot \sin(b),
\end{equation}
where $l$ and $b$ are the galactic longitude and latitude.
In our analysis, we consider only the 2019 pixels whose galactic
latitude is 20 degrees or above.
 The COBE 53 GHz (A+B)/2 and (A-B)/2 are analyzed and the reduced three-point
 functions are shown in Fig. 7a, 7b. The result is consistent
with Gaussian fluctuations and there is no deviation above the statistical
uncertainty. We conclude that the current analysis shows that the
statistics of CBR at the COBE scale are probably Gaussian.

\section{Statistical Tests on Intermediate Angular Scales ($\theta \sim
1^{\circ}$)}

The Gaussianity question is hard to resolve on the COBE scale ($
\theta \sim 7^{\circ}$) even if there is no noise in the experimental
data. One
 should expect the CBR on the large angular scale to be Gaussian simply by
virtue of the central limit theorem \cite{scherrer}. Furthermore, there is
intrinsic uncertainty in the statistical
quantities measured  in our local universe  due to the
cosmic variance \cite{cvar}, which makes
it harder to discriminate between Gaussian and non-Gaussian
 fluctuations on the
 COBE scale alone.
On intermediate angular scales ($\theta \sim 1^{\circ}$),
the cosmic variance is small
and the chance to detect a non-Gaussian signal is  higher. As the
  data on these scales is accumulating, it is timely to consider seriously
testing Gaussianity on these scales, since data in some experiments
\cite{cheng}\cite{south} show clear non-Gaussian features. Although the current
experiments
are inclusive, due to the possible foreground contamination, it gives us hope
that the Gaussianity problem will be resolved experimentally in the near
future.
The EP characteristic and the three point correlation will apply equally on
both large and small angular scales if the sky coverage of the experiment
is substantial. However, as the current  state-of-the-art intermediate scale
CBR experiments cover a tiny fraction of the sky, more sophisticated
statistical tools \cite{neil} are required to carry out the tests.
In this section, we will introduce and discuss the bispectral analysis and
  the Hotelling's $T^{2}$
 statistic. We will
 show that  the $T^{2}$ statistic is a powerful statistical quantity to use on
these scales and we also estimate the minimun data sample size to carry out
the Gaussian test through $T^{2}$ statistic.

In most current intermediate scale experiments, the data is sampled either in
thin long strips or an annulus around an axis. In both cases, the data is one
dimensional.  The three-point function and bispectral analysis of 1-d
data are well studied by statisticians and much of the specific techniques
and mathematical details of this section are contained in monographs by
\cite{rao}\cite{brillinger} which
 interested readers should consult. One can arrange the dataset as a time
series, $X_{t}$, where in the present case, ``time'' $t$ refer to successive
positions
in the sky. The data is usually edited so that the mean is removed.
In this case, the three-point function is simply
\begin{equation}
\xi_{t_{1}, t_{2}} = <X_{t} X_{t+t_{1}} X_{t +t_{2}}>_{t}
= {1\over N} \sum_{i=1}^{N} X_{i} X_{i + t_{1}} X_{ i + t_{2}},
\end{equation}
where the  expression on the left-hand side
 is an estimate of three point function from a data sample
and $N$ is the size of the sample.
Since the temperature anisotropies are always real-valued and assumed to
be homogeneous and isotropic, the three point function has the following
symmetry:
\begin{equation}
\xi_{t_{1}, t_{2}} = \xi_{t_{2}, t_{1}}
= \xi_{-t_{1},  t_{2} -t_{1}}  = \xi_{t_{1} -t_{2}, -t_{2}}
\end{equation}
The bispectral density $f(\omega_{1}, \omega_{2})$ is the Fourier
transformation of $\xi_{t_{1}, t_{2}}$, where
\begin{equation}
f(\omega_{1}, \omega_{2}) = {1\over (2\pi)^{2}} \int dt_{1} dt_{2}
e^{-it_{1}\omega_{1} -it_{2}\omega_{2}} \xi_{t_{1}, t_{2}}; \ \
-\pi\leq\omega_{1}, \omega_{2}\leq \pi.
\end{equation}
In general, $f(\omega_{1}, \omega_{2})$ is a complex function,
following the symmetry of $\xi$, one obtains the following symmetry relation
for
$f$:
\begin{equation}
f(\omega_{1}, \omega_{2}) = f(\omega_{1}, \omega_{1}- \omega_{2})
=f(-\omega_{1}- \omega_{2}, \omega_{2}) =f^{*}(-\omega_{1}, -\omega_{2})
\end{equation}
Because of the symmetry, one just has to estimate the bispectral
density in a small portion of $(\omega_{1}, \omega_{2})$
parameter space.

The unique feature of the Gaussian process is  that the bispectral
density vanishes for all $\omega$, i.e.,
$f(\omega_{1}, \omega_{2}) = 0$ for all $\omega_{1}, \omega_{2}$.
To test this hypothesis, one can use the Hotelling's $ T^{2}$
 statistics\cite{rao}, which is  constructed from the
bispectral estimates defined on a ``coarse-grained'' frequency grid,
$(\omega_{i}, \omega_{j})$, where
\begin{equation}
\omega_{i} = {i\pi\over K}, \omega_{j} =
{j\pi\over K}, i =1, 2, ..., L; \ \ j =i+1, ..., \gamma (i),
\end{equation}
where $L = [{2K\over 3}]$, $\gamma(i) = K - [{i\over 2}] -1$.
The parameter $K$ is chosen to be much smaller than the sample size
so that the frequency grid is ``coarse-grained''.
Let $\eta_{ij} = f(\omega_{i}, \omega_{j})$ and rearrange $\eta_{ij}$ into
a vector ${\bf{\eta}} = (\chi_{1}, ..., \chi_{P})$, where $P = \sum_{i =1}^{L}
(\gamma(i) -i)$ so that for each $l, 1\leq l \leq P, \ \ \chi_{l} = \eta_{ij}$.

 To estimate the  bispectral density at each ``coarse'' grid point, one can
construct a ``fine'' frequency grid around each $(\omega_{i}, \omega_{j})$
point. Specifically,
\begin{eqnarray}
\omega_{i_{p}} = \omega_{i} + {p D\pi\over N}, p =-r, -r +1, ..., r
\nonumber
\\
\omega_{i_{q}} = \omega_{i} + {q D\pi\over N}, q =-r, -r +1, ..., r, (q\neq 0)
\end{eqnarray}
where the distance $D$ is chosen so that the bispectral estimates at
neighboring points on this fine grid are approximately uncorrelated.
To insure that points in different ``fine'' grids don't overlap, it is required
that $D \leq {N \over K (2 r +1)}$.  Since the total number of points in each
``fine'' grid is $(4 r +1)$ and there are $K^{2}/3$ ``coarse-grained'' grids,
thus the constraint on parameter $r$ is: $(4 r +1) K^{2}/3 < N$.

 Let
$\hat{f} (\omega_{i_{p}}, \omega_{i_{q}})$ denote the estimated bispectral
density function at the points $(\omega_{i_{p}}, \omega_{i_{q}})$. Due to
the careful choice of grid point, one may regard the set of estimators
$\{\hat{f} (\omega_{i_{p}}, \omega_{i_{q}})\}$ as $n = 4 r +1$ uncorrelated
and unbiased estimates of $f (\omega_{i}, \omega_{j})$. Forming the bispectral
estimates $\hat{f} (\omega_{i_{p}}, \omega_{i_{q}})$ into a $n$ column vector,
denoted by ${\bf \xi} = ( \xi_{1}, \xi_{2}, ..., \xi_{n})$, then, at each
``coarse-grained'' grid point $\chi_{l}$, there will be an estimated bispectral
density $\xi_{i}^{(l)}$ from the ``fine-grid''.  When the
sample size $N$ is large,
 $\xi_{i}^{(l)} (i =1, 2, ..., n), $ is distributed as complex normal
with mean $\bf \eta$ and covariance matrix $\Sigma_{\xi}$. The maximum
likelihood estimates of $\bf \eta$ and $\Sigma_{\xi}$ are:
\begin{equation}
\hat{\bf \eta} = {1\over n} \sum_{i=1}^{n} \xi_{i}^{(l)},
\ \ \hat{\Sigma} = { A\over n}, \ \ A_{ij} = \sum_{k=1}^{n} ((\xi_{ik} -
{\bf \eta}) (\xi_{jk}-{\bf \eta})^{*}).
\end{equation}
The Hotelling's $T^{2}$ statistics are defined as:
\begin{equation}
T^{2} = n \hat{\bf\eta} A^{-1} \hat{\bf \eta}.
\end{equation}
Under the Gaussian assumption, the mean vector $\eta =0$ and
the statistic
\begin{equation}
F = {2 (n-P)\over 2 P} T^{2}
\end{equation}
is distributed as a central F-distribution with (2P, 2(n-P)) degrees of
freedom.

To test Gaussianity through degree scale experiments, the sample size
should be large enough to avoid ``sample variance'' \cite{scott}.
The minimal sample size that could be used to carry out the Gaussian
statistical tests can be estimated as follows: to carry out the
 bispectral analysis and Hotelling's $T^{2}$
statistics, one should use at least two
``coarse'' grids,  and  because the correlation angle of the beam-smoothed
CBR anisotropies is around $ \theta_{*}$, where $\theta_{*} \simeq
 1.8 \theta_{s}$ for Harrison-Zeldovich primordial spectrum \cite{be}. Thus, we
need at least two ``fine'' grids on either side of the ``coarse''
grid. The minimum parameters to test Gaussianity is
 $L\geq 2$ and $r \geq 2$, which in turn gives $ K \geq 4$. Thus,
the minimum sample size is: $N \geq ( 4 r +1) K^{2}/3 =48$. This is about
twice the current largest sample at the half degree scale \cite{wollack}.
This result is encouraging: it suggests that we don't need a full sky coverage
with half-degree resolution to carry out the Gaussianity test, and the
Gaussianity question can be resolved by ground-based experiments or balloon
flights.
 With  regard to  sampling
in the sky, the data should sample as sparsely as possible to avoid possible
correlation between different data points, and it will be better to sample in
thin long stripes or annuli to take advantage of the simplicity of one
dimensional
dataset.  Currently, most degree scale
experiments have fewer data points and are thus inadequate to perform
bispectral
analysis. However, we expect the situation will change dramatically soon,
and the results we discuss here will be helpful for designing future
experiments.

\section{Gaussianity of CBR on Small Scales}

 In this section, We will show how to use the $T^{2}$ statistic
 by analyzing the recent
96 point RING data from OVRO \cite{meyers}.
 A special class of non-Gaussian signal,
point sources, which are interesting in their own right,
 is separated out and discussed in detail in the next section.
 The
data sample is weighted according to the error in each pixel. The sample
size $N=96$. The estimate of the
three-point function is shown is Fig. 8a and the
the real and imaginary part the bispectrum is shown in Fig. 8b, 8c.
We have used an optimum window function (see \cite{rao} for details)
to smooth the discrete data. For Gaussian distributed data, both the
three-point
 function and bispectrum should vanish. The results shown in Fig. 8a, 8b, 8c
 already suggest that the data might be non-Gaussianly distributed.
To show how statistically significant
 the deviation from Gaussian distribution,
we can use the Hotelling's $T^{2}$ statistics. The parameters we choose for
RING data are:
$ K =4, r =2, n =9, d = 4, P=2, L=2$. With this choice of parameters,
the statistical distribution for $F$ is shown in Fig. 9. The 95\% confidence
level (C.L.) upper limit of $F$ is $F_{c} = 3.15$ if the data are Gaussian
distributed.
Hotelling's $T^{2}$ statistic
estimated from RING data  is $T^{2}=5.76$, or $F = 7 T^{2}/2 = 20.1$ which is
much larger than the 95\% C.L. upper limits for Gaussian distributions.
Thus, we conclude that the data are probably not consistent
with Gaussian statistics.

Even though  the data  failed the statistical
tests we proposed, one cannot conclude that the
non-Gaussianity is due to non-trivial interactions in the inflationary
cosmology or the topological defects produced in the early universe.
The non-linear gravitational evolution will produce non-Gaussian signals
which have to be carefully studied and subtracted to gain some  knowledge
about the Gaussianity of primordial perturbations.  Part of the
answer to this important issue is contained in the next section and we will
not discuss here.

\section{Point Sources}
On intermediate angular scales ($ \sim 1^{\circ}$), the current datasets are
too small to carry out the  three-point correlation and the angular bispectrum
analyses we developed in the previous sections.
 To test Gaussianity based on the
small dataset available (usually about 10 data points), the statistics
have to be very custom-designed to be useful \cite{neil}. However,
a clear non-Gaussian signature will be the point-like CBR  anisotropies.
In fact, two candidates of such sources are detected in the MSAM experiment
\cite{cheng}. One source, located in a dust-free region, has a flux of
$3.7\pm 0.9 \Jy$ at $5.6 {\rm cm^{-1}}$. Another candidate, has a flux
 of $2.9 \pm 0.7
{\rm Jy}$,
is located $4.3^{\circ}$ away from the first sources. Both sources
are compact and have angular size less than
the beam width $\sigma = 0.425 \theta_{FWHM} = 12^{'}$. Assuming that the
angular size of the sources are half the beam width, one can find the the flux
intensities are $(3.8\pm 0.8) \times 10^{5} {\rm Jy Sr^{-1}}$ and $(3 \pm 0.6)
 \times 10^{5} {\rm Jy Sr^{-1}}$. As a comparison, the flux of a $ \delta T$
temperature fluctuation will produce a flux
\begin{equation}
I_{\nu} = B_{\nu}(T_{0}) {\exp(h\nu/kT_{0}) \over \exp(h\nu/kT_{0}) -1}
({h\nu\over kT_{0}}) { \delta T\over T_{0}},
\end{equation}
where $T_{0} = 2.73 K$ is the CBR temperature and  $B_{\nu}$ is the CBR
flux at frequency $\nu$. At $5.6 \ \ {\rm cm^{-1}}$, the CBR flux is
$B_{\nu} = 1.5 \times 10^{9} {\rm Jy Sr^{-1}}$. Thus, for a temperature
anisotropy of $40 \mu K$, which is the theoretically estimated temperature
anisotropy at the half degree angular scales for CDM with standard
recombination \cite{gorski}, the expected flux
is
\begin{equation}
I_{\nu} = 6.8 \times 10^{4} ({\delta T \over 40 \mu K}) {\rm Jy Sr^{-1}}.
\end{equation}
Thus, the first sources correspond to a $5 \sigma$ and the second sources
correspond to a $4 \sigma$ peak.

 In the Gaussian picture, for a $\nu -$peak,
the mean size $\bar{\theta}$ and the average distance between peaks $\bar{d}$
 are  given by \cite{be}:
\begin{equation}
\bar{\theta} \sim \sqrt{2} {\theta_{*}\over \gamma \nu} \sqrt{1 -{1\over
\nu^{2}}}, \ \ {\rm for} \ \  \nu \gg 1,
\end{equation}
\begin{equation}
\bar{d} = {2/3} (\pi n_{\nu})^{-1/2},
\end{equation}
where $\theta_{*} = 1.2 \theta_{s}$ and $n_{\nu}$ is the number  density of
$\nu$ peaks. For large $\nu$,
$n_{\nu}$ is given by:
\begin{equation}
n_{\nu} = {\gamma^{2} (\nu^{2} -1) \over (2\pi)^{3/2}} \exp (-\nu^{2}/2).
\end{equation}
Thus, for $4 \sigma$ peaks, the mean size will be $0.71 \theta_{*}$,
which is marginally consistent with observation, but the
 the mean distance between rare peaks will be: $ \sim 50 \theta_{s} \sim
20^{\circ}$, which is much larger than the
angular separation between the sources.
One may try to  explain both point sources by Gaussian statistics,
 assuming that they are just 3 $\sigma$ peaks and fit the low limit of the
observed flux.
 Then the average distance between peaks
is  $ 11 \theta_{s}$, which is roughly the same as the observed value. However,
in this case, the averaged angular sizes  of the peaks are $1.2 \theta_{s}$,
which is larger than the beam width. We conclude that if these sources
 are of CBR origin, they are not consistent with Gaussian statistics.

Various topological defects, notably  soft domain wall bubbles \cite{bubble},
the global
monopoles \cite{monopole}  or textures \cite{sigma},
 are capable of producing
spot-like CBR anisotropies of any size by appropriately choosing model
parameters. However, before one relies on topological defects as an answer, one
has to
 filter out the foreground contaminations carefully.
Several types of radiation may contribute the point-like sources observed
in the experiments. One of them is
 the Sunyaev-Zeldovich (SZ) effect from  rich
 clusters. The scattering of microwave  photons by hot
electrons in the intracluster gas will make a cluster a powerful source of
submillimeter radiation. The typical angular size of the core of the hot gas
is of order arcminutes,
 and the flux density is given by \cite{sz}:
\begin{equation}
F_{\nu} = y (x {e^{x} +1\over e^{x} -1} -4) \cdot {xe^{x} \over e^{x}-1} \cdot
B_{\nu} (T_{0}),
\end{equation}
where $x = {h\nu\over kT_{0}}$ and $y = \int {kT_{e}\over m_{e}c^{2}}
\sigma_{T}
n_{e} dl$. For a typical rich cluster, $T_{e} \sim 10^{7} K$,
$n_{e} \sim 10^{-3} {\rm cm^{-3}}$ and $l \sim 1 {\rm Mpc}$, the estimated
$y$ parameter is around $10^{-4} - 10^{-5}$, The flux density
is around $3 \times 10^{5} {\rm Jy Sr^{-1}}$ which is exactly what the MSAM
experiment
observed. Thus, it is very likely that the observed sources are due to SZ
effects of unresolved rich clusters in the field.

 The multifrequency channel methods is widely used to separate the local
contamination from true CBR signals. The method may not be effective
to single out the SZ effect. As we showed in Fig. 10,  the spectral
index of the SZ effect is very close to that of primordial temperature
fluctuations at low frequencies ($\nu < 100$ GHz). When the frequency gets
higher ($\nu > 200$ GHz), there is a small deviation, but at this frequency
range, the dust emission will dominate. A possible way to discriminate
the SZ effect is through polarization of  the radiation from candidate point
sources. Due to the peculiar velocity of the rich clusters, the radiation
will be polarized in SZ effect \cite{sz}. However,
the point-like CBR anisotropies from topological defects will not.
But since  it is currently hard to measure  the polarization
of the radiation down to required accuracy, we won't  discuss
 this approach in detail  here.

 Apart from the SZ effect,  primeval dust \cite{bond} or a population of
IRAS-like galaxies at high
redshift \cite{bh}
may have substantial contributions at the submillimeter range. For dust grains,
if one assumes the emissivity of the dust is $\sim \nu^{\alpha}, \alpha \approx
1.5$, the flux spectrum of dust emission is given by:
\begin{equation}
f_{\nu} \sim ({h\nu\over kT_{d}})^{4 +\alpha} {1\over \nu} {1\over e^{
{h\nu\over kT_{d}}} -1}.
\label{cold}
\end{equation}
The peak of the distribution is located at
\begin{equation}
\nu_{pk} = 4.5 kT_{d} /h = 3750 {\rm GHz} (T_{d}/40K).
\end {equation}
Thus, in order for the peak of a hot dust spectrum $(T_{d} = 40 K)$ to be
 redshifted
into the 300 GHz range which the MSAM experiments operates, the redshift
of the epoch of  formation of primeval galaxies
 should be around $z \sim 10$.  The angular size of the dust envelope is
\begin{equation}
\theta = {l (1 +z) \over D}, \ \ D = 2 H_{0}^{-1} ( 1- {1\over \sqrt{1 +z}}),
\end{equation}
where $l$ is the proper size of the dust envelope, which is about $10 - 100$
 kpc. Thus the typical angular size of the possible point sources produced by
the
primeval galaxies is
\begin{equation}
\theta  = 0.03^{'} {l\over 10 {\rm kpc}} \ \ {1+z\over 10} \ \ ({h\over
0.5})^{2}.
\end{equation}
The observed flux density is
\begin{equation}
S_{\nu} = {L(\nu_{1}) \over 4 \pi D^{2} (1 + z)}, \nu_{1} = (1+z) \nu.
\end{equation}
Assuming that,  in the rest frame of the sources, the intrinsic flux
is peaked around 3000 GHz and the luminosity is $L$, then  the flux
density is
\begin{equation}
S_{\nu} = 10^{-3} {\rm Jy Sr^{-1}}
 ({L\over 10^{13} L_{\odot}}) ({10\over 1 +z})
({3000 {\rm GHz}\over \nu})
\end{equation}
Both the angular size and the flux density are too small to account for the
observed flux in the MSAM experiment \cite{cheng}. Thus we can conclude that
primeval
dust or the distant infrared
 galaxies do not account for the point-like sources observed.

As we showed in Fig. 11, which is plotted according to  Eq. (\ref{cold}),
 it is clear that if there is a population
of cold dust ($T_{d} \sim 4K$), then the flux density will peak near where
the experiment operates. A uniform background of such cold dust is
impossible unless the density the low enough so that the optical light from
distant quasars will not be absorbed. However, clumpy cold dust is helpful
to explain the experimentally observed point-like sources.
A possible scenario to explain the
spot-sources based on cold dust is the following: there is a population of
very quiet galaxies where most of the star formation activities are shut
down, so that there will be very low intensity radiation in the far-infrared
regime. Thus, this population of galaxies is not observed by the IRAS flux
limited survey. However,  as the hot dust cools down to around 4K, they
become powerful submillimeter emitters. The spectrum of cold dust emission
is shown Fig.  11. Multi-spectral analyses can shed light on the possible
spectral parameter of the sources. The problem with the cold dust scenario
is that as one look back in time, these sources used to be very powerful
infrared sources  because the radiated flux $\propto T_{d}^{4+\alpha}$.
 Due to the abundance one observes today, we can estimate
the luminosity at high redshift, which is much brighter. However,
the newest result from the COBE FIRAS \cite{mather}
has already put a stringent limit on the possible
evolution of infrared galaxy luminosity function. Detailed modelling is in
progress.
But we are pessimistic about explaining the point-sources based on cold
dust scenarios.

In conclusion, the point-like CBR anisotropy is a clear non-Gaussian signature.
If the future studies confirm that the point sources candidates are truly
CBR fluctuations, it will be an exciting new chapter. It will provide direct
evidence that topological remnants left over from the early universe do exist.

\section{Conclusion}

In summary, the following points which are related to the test of the
Gaussian nature of the primordial fluctuations are discussed in this paper:

\noindent
(1) We listed the skewness and kurtosis in various physically motivated
models, with and without noise. We also discussed the use of the multivariant
skewness and kurtosis to quantify the deviation of a distribution
from Gaussian.

\noindent
(2) We discussed in detail the  Euler-Poincare characteristic of random
fields. We showed that the Euler-Poincare characteristic will not be
a good discriminator between Gaussian and non-Gaussian random field
 when the noise
is comparable to the signal.

\noindent
(3) We stress the use of the three-point temperature correlation function to
test Gaussianity. The predictions from various models are discussed and
the COBE 53 Ghz data is analyzed. The analysis shows that the fluctuations
are probably Gaussian on the COBE scale.

\noindent
(4) We discussed the detailed statistical tests on intermediate
 angular scales.
The bispectral analysis and Hotelling's $T^{2}$ statistics are emphasized.
We also
discussed briefly the sampling technique and minimum sample size to test
Gaussianity statistically  on half degree scales.

\noindent
(5) We discussed testing Gaussianity on small angular scales (arcminute
scales). The RING data from OVRO is analyzed and shown to be propably
non-Gaussian.

\noindent
(6) We discussed looking for  point-like sources as a way to test
Gaussianity and hunt for topological defects in small scale CBR experiments.
 The  SZ effects and the effects of  primeval dusts  are discussed.

The current status of testing Gaussian nature of CBR anisotropies
are summarized in Table 3.

\vskip 0.3 in

\centerline{\bf Acknowledgments}
\medskip
I want to thank my advisor,  David N. Schramm, for his constant encouragement
and guidance over the years. I want to thank  Stephan Meyer,
Albert Stebbins, Neil Turok, George Smoot and Lyman Page for
 very helpful discussions. I also want to thank Roberta Bernstein
and Brian Fields for reading and editing the paper.
This work is support in part by NSF grant \# 90-2269
and by NASA grant \# NAGW 1321 at the University of Chicago and by DoE and by
NASA through
grant \# NAGW 2381 at Fermilab.
The COBE datasets were developed by the NASA
         Goddard Space Flight Center under the guidance
         of the COBE Science Working Group and were
         provided by the NSSDC.

\newpage

\appendix

\centerline{\Large\bf Appendix}

\medskip

\section{Gaussian Nature  of Perturbations from Inflation}

  In this appendix, we will discuss the Gaussianity of the
primordial fluctuations produced in  inflation. We pick one
simple model, Linde's chaotic inflation model \cite{linde} to analyze.
The approach is rather heuristic but brings into focus the Gaussianity
problem. For more rigorous treatment, see \cite{salopek}.

 The basics of the inflationary dynamics are the following:
 there
exists an epoch where the universe is dominated by the
vacuum energy of a scalar field ${\phi}$.  The Friedman equation
which describes the evolution of the background metric is:
\begin{equation}
H^{2} = {8\pi G \over 3}  \rho,  \ \  \rho = 1/2 {\dot{\phi}}^{2}
+ V(\phi),
\end{equation}
and the dynamical evolution of $\phi$ field is:
\begin{equation}
\ddot{\phi} + 3 H \dot{\phi} + {\partial V(\phi)\over \partial\phi} = 0,
\end{equation}
where the inflaton potential for chaotic inflation is simply
\begin{equation}
 V( \phi ) = \lambda \phi^{4}.
\end{equation}
The model is easy to analyze in the slow-roll regions where
$\ddot {\phi} \ll H \dot{\phi}$, which is sastified for
$\phi > m_{pl}/\sqrt{2\pi}$. By using the gauge-invariant
parameter $\zeta$ introduced by \cite{bardeen}, where
\begin{equation}
\zeta_{k} \approx {H (\Delta\phi) \over \dot{\phi}},
\end{equation}
$\Delta \phi$ is the zero-point quantum fluctuation of the
field $\phi$ along its classical trajectory. The probability
density function for $\Delta \phi$ is a Gaussian with zero mean
and variance given by $(\Delta\phi)^{2} = ({H \over 2\pi})^{2}$.
Let us denote $ {2\pi \Delta \phi\over H}$ by $x$, then $x$ is a random
variable with standard normal distribution.

Expressed in terms of $x$,
 $\zeta$ is given by:
 \begin{equation}
{\zeta_{k}} = {H^{2} \over {\dot{\phi}}} \cdot x
\end{equation}
In the slow approximation, the quantum fluctuation of $\dot{\phi}$
is negligible. Thus, if the Hubble parameter is a constant during
inflation (which is the case for original exponential inflation),
 one can conclude that the primordial density fluatuation
is Gaussian. However, since $H$ is related to $\phi$ locally, the fluatuation
in $\phi$ will give rise to a fluctuation in $H$. Taking this into account,
\begin{equation}
\zeta_{k} = {H_{c}^{2}\over\dot{\phi}}  [ x + \sqrt{\lambda \over 12 \pi}
{\phi \over m_{pl}} x^{2} ],
\end{equation}
 where $H_{c}$ is the classical value and $m_{pl} =
{1\over \sqrt{G}}$ is the Planck mass. The usefulness of the $\zeta$
parameter lies in the fact that it is a constant throughout the
inflationary, radiation and matter dominated epoch,  and the
Sachs-Wolfe contribution to the CBR temperature fluctuation
is given by \cite{salopek}:
\begin{equation}
{\delta T \over T} \simeq  {\zeta \over 5}
\end{equation}
Thus, the large angular temperature anisotropy is given by:
\begin{equation}
{\delta T \over T} \simeq {H_{c}^{2}\over 5 \dot{\phi}}  [ x + \sqrt{\lambda
\over 12 \pi} {\phi \over m_{pl}} x^{2} ]
\end{equation}
The statistics of CBR fluctuations are non-Gaussian because of the
$x^{2}$ term. This expression for the deviation from
Gaussianity is rather generic in inflationary models and will lead to
the functional form of the three point correlation we will give  in section 4.
The difference is
that various models will produce different skewness. In the one-field
chaotic inflation model we treat now, the skewness is given by:
\begin{equation}
\mu_{3} = 3 \sqrt{\lambda \over 12 \pi} {\phi \over m_{pl}}.
\end{equation}
During the slow roll \cite{kolb},
\begin{equation}
{H^{2}\over \dot{\phi}} = 4 \sqrt{2/3}\lambda^{1/2} N_{e}^{3/2},
\end{equation}
where $N_{e} \sim 60$ is the number of $e-$folds the scale factor inflates
during  inflation. In order for the amplitude of the fluctuation
to be of the same order of magnitude as the COBE observation \cite{cobe},
the self-coupling constant $\lambda$ of the inflaton field is given by
$\lambda \sim 10^{-15}$. For such a weakly coupled field, the deviation
from Gaussian is estimated to be:
$\mu_{3} \sim 10^{-7}$
with $\phi \sim \phi_{i} \sim 5 m_{pl}$, which is negligibly small.

In conclusion, we have shown that the primordial fluctuation
from slow roll inflation is very close to a Gaussian.
The deviation from Gaussianity is of order $10^{-7}$ in the chaotic inflation
model, and   the result is  general to other one-field inflation models.
The root cause is  the small coupling constant of the inflaton
potential, which is required to give the right amplitude for primordial
fluctuations. Thus,  Gaussian-distributed CBR anisotropies are a nature
result of lots of  inflation models.

\newpage
\centerline{\bf Table Captions}
\vskip 0.2 in

\noindent
Table 1: The skewness and kurtosis of the temperature
anisotropies in various non-Gaussian models.

\noindent
Table 2: Testing the properites of signal and noise in COBE
53 GHz map. $W_{n}$ is the statistic for testing mutual independence
of noise. Under the null hypothesis that the noise is uncorrelated,
$W_{n}$ will distribute as a standard normal. $W_{c}$ is for testing
correlation between signal and noise, and $W_{s}$ is for testing the
correlation between signals. Under the null hypothesis (signal and
noise are not correlated; signals are not correlated), both
$W_{c}$ and $W_{s}$ are distributed as standard normals.

\noindent
Table 3: A summary of current status on testing Gaussian nature of
CBR temperature anisotropies on different angular scales.

\newpage
\vskip 0.1 in
\centerline{Table 1}
\vskip 0.2 in
\begin{tabular} {|c|c| c|c| c |}
\hline

 &
\multicolumn{2} {c|} {No Noise} &
\multicolumn{2}{c|} {Signal to Noise Ratio 1:1} \\ \cline{2-5}
 Non-Gaussian Sources &Skewness & Kurtosis &Skewness & Kurtosis \\ \hline
Soft Domain Wall\footnotemark[1] &1.4 & 3.75 &0.5 &0.94 \\ \hline
Cosmic string\footnotemark[2] & 0 &1.5 &0 &0.38 \\ \hline
Global monopole\footnotemark[3] &0.82&1.25 &0.29 &0.31 \\ \hline
Global Texture\footnotemark[3] &0.71 &0.94 &0.25 &0.25 \\ \hline
$O(N)$ $\sigma$-model\footnotemark[3] &$\sqrt{2/N}$ &15/4$N$ &$\sqrt{1/4N}$
&15/16$N$ \\ \hline
SZ from Rich Cluster\footnotemark[4] & -0.66  & 2.72  & -.28 &0.22  \\ \hline
\end{tabular}
\footnotetext[1]{Extrapolated from $O(N)$ model}

\footnotetext[2]{On Scales of several arcminutes}
\footnotetext[3]{COBE sensitive scale, no beam-smoothing}
\footnotetext[4]{On Scales of several arcseconds to arcminutes}

\newpage
\vskip 0.2 in
\centerline{Table 2}
\vskip 0.2 in

\begin{center}
\begin{tabular} {|c| c|c| c |}
\hline
 & estimate & Confidence Level \\ \hline
$W_{n}$ & 2.54 & $1.58\times 10^{-2}$ \\ \hline
$W_{c}$ & 1.21 & 0.19 \\ \hline
$W_{s}$ & 8.73&$1.1\times 10^{-17}$ \\ \hline
\end{tabular}
\end{center}
\vskip 0.2 in

\newpage
\vskip 0.2 in
\centerline{Table 3}
\vskip 0.2 in
\begin{tabular} {| c|c|c| c |}
\hline
 Angular scale&Small  &Intermediate
& Large \\
 $\theta_{s}$ &($\sim$ arcminutes) & ($\sim 1^{\circ}$) & ($\gg 2^{\circ}$) \\
\hline
 & & South Pole, & \\
 Current&  &MAX, MSAM, & COBE \\
Experiments&OVRO &   SK93, PATHON, & MIT balloon   \\
 &  & White Dish,...& \\ \hline
 & & Inflation: & \\
 Theoretical& & Gaussian;  & \\
expectations&Non-Gaussian &  Defects: &Gaussian \\
 & &  non-Gaussian. & \\  \hline
Analysis & & Sky coverage & \\
 from &Non-Gaussian & is still  & Gaussian \\
 experiments&&  too small & \\ \hline
 &Source & Eventually & Central limit theorem; \\
Comments& contaminations &  Decisive & Cosmic variance \\ \hline
\end{tabular}
\vskip 0.2 in

\newpage
\centerline{\bf Figure Captions}

\noindent
Fig. 1: The probability distribution functions of the modified
$\chi^{2}$ distribution with $n$ degrees of freedom.
 The dotted line is for $n$=4, the short dash line is for $n$=8,
the lone dash line is for $n=16$ and the solid line is the standard
normal distribution. The values for $n$ are chosen because the
corresponding distributions
describe the domain wall, global string and texture, respectively.

\noindent
Fig. 2: The simulated unsmoothed Gaussian noise $\delta$ in a $10^{\circ}
\times 10^{\circ}$ patch. The solid line is for the contour
$\delta =0$, the dotted line is for the contour $\delta = \sigma$
and the short-dashed line is for $\delta = 2 \sigma$, where $\sigma$ is the
standard deviation of the noise.

\noindent
Fig. 3: The statistic of EP characteristic for a Gaussian
random field. The soild line is the mean EP characteristic
and the dotted line is the $1 \sigma$ uncertainty estimated
from Eq. (\ref{ep}). The unit of the vertical axis is arbitary.

\noindent
Fig. 4a: The contour plot of the sky signals $\delta$ if they are dominated by
a  quadrupole. The solid line is for the contour
$\delta =0$, the dotted line is for the contour $\delta = \pm \sigma$
and the short-dashed line is for $\delta = \pm 2 \sigma$, where $\sigma$ is the
standard deviation of the observed sky signals.

\noindent
Fig. 4b: The contour plot of the sky signal $\delta$
 when the signal to noise ratio is 1:1, where
the signal is a  quadrupole. The solid line is for the contour
$\delta =0$, the dotted line is for the contour $\delta = \pm \sigma$
and the short-dashed line is for $\delta = \pm 2 \sigma$, where $\sigma$ is the
standard deviation of the observed sky signals.

\noindent
Fig. 5: The genus curve for $\chi^{2}$ distributions with $n$
degrees of freedom. The solid line is for the random Gaussian
field, the dotted line is for $n=12$ and the dashed line is
for $n=21$, which describes the noisy case.

\noindent
Fig. 6: The theoretical prediction of reduced three-point function in three
different models: The dotted line is for inflation, the short dash line
is for $O(N)  \ \ \sigma$ model and the long dash line is for LTPT. The reduced
two point function, which is the solid line, is modeled as $\exp(-\theta^{2}
/\theta_{c}^{2})$, where $\theta_{c} = 13.5^{\circ}$.

\noindent
Fig. 7a: The reduced three point-functions
estimated from COBE DMR 53 GHz (A+B)/2 map and (A-B)/2 map.
The solid line is for the (A+B)/2 map and the dotted line is for
the (A-B)/2 map.

\noindent
Fig. 7b: The estimate and error of the three point function
for COBE 53 GHz (A+B)/2 map. The result is consistent with
prediction from a Gaussian field (which is zero).

\noindent
Fig. 8a: The reduced three-point function estimated from  RING data.
We place 96 RING data points on a one dimensional lattice and
the horizontal axis is the number of lattice spacing between
the points used for estimating the reduced three point function.

\noindent
Fig. 8b: The plot of the
 real part of the estimated bispectrum $f(\omega_{1}, \omega_{2})$
estimated from the  RING data. In this plot, we choose
$\omega_{1} = \omega_{2} =\omega$.
 The  frequencies  $\omega$ is plotted
 in unit of $\pi/20$.

\noindent
Fig. 8c: The imaginary part of the estimated bispectrum $f(\omega_{1},
\omega_{2})$  from the  RING data. In this plot, we choose
 $\omega_{1} = \omega_{2} = \omega$.  The  frequency  $\omega$ is
plotted  in units of $\pi/20$.

\noindent
Fig. 9: The statistical distribution  of $F = {2(n-P) \over 2 P} T^{2}$.
We choose $P =2$ and $n =9$ for the RING data.

\noindent
Fig. 10: The frequency dependence of the antenna temperature
for CBR and for SZ effect. The soild line is for CBR and the dotted
line is for SZ. The antenna temperature is normalized so that it is
unity at low frequencies.

\noindent
Fig. 11: The flux density of the 4K cold dust. The emissivity
of the dust is chosen to be $\approx \nu^{1.5}$. The unit of the vertical
axis is arbitary.

\end{document}